\documentclass[useAMS]{mn2e}
\usepackage{psfig}
\newif\ifAMStwofonts

\def\be{\begin{eqnarray}}
\def\ee{\end{eqnarray}}
\def\beq{\begin{equation}}
\def\eeq{\end{equation}}

\def\etal{{\it et al.}}

\def\HI{{\hbox{H~$\scriptstyle\rm I\ $}}}
\def\HII{\hbox{H~$\scriptstyle\rm II\ $}}

\def\MgII{\hbox{Mg~$\scriptstyle\rm II\ $}}

\def\nHI{{\rm HI}}

\def\lya{Ly${\alpha}\,\,$}

\def\lesssim{\mathrel{\hbox{\rlap{\hbox{\lower4pt\hbox{$\sim$}}}\hbox{$<$}}}}
\def\gtrsim{\mathrel{\hbox{\rlap{\hbox{\lower4pt\hbox{$\sim$}}}\hbox{$>$}}}}

\def\gtsima{$\; \buildrel \over \sim \;$}
\def\ltsima{$\; \buildrel < \over \sim \;$}
\def\prosima{$\; \buildrel \propto \over \sim \;$}
\def\gsim{\lower.5ex\hbox{\gtsima}}
\def\lsim{\lower.5ex\hbox{\ltsima}}
\def\simgt{\lower.5ex\hbox{\gtsima}}
\def\simlt{\lower.5ex\hbox{\ltsima}}

\def\simpr{\lower.5ex\hbox{\prosima}}

\def\etal{{\frenchspacing et al. }}
\def\ie{{\frenchspacing\it i.e. }}
\def\eg{{\frenchspacing\it e.g. }}

\def\be{\begin{eqnarray}}
\def\ee{\end{eqnarray}}

\def\xhi{${x}_{\rm HI}$}

\def\rmax{$R^{max}_{w}$ }
\def\rw{$R_w$}

\def\lya{Ly$\alpha$ }

\title[Transmission windows of quasars]
{Interpreting the Transmission Windows of Distant Quasars}

\author[Maselli, Ferrara \& Gallerani]
{A. Maselli$^{1}$, A. Ferrara$^{2}$, S. Gallerani$^{3,4}$\\\\
$^1$ Max-Planck-Institut f\"ur Astrophysik, Karl-Schwarzschild Str. 1, 85748 Garching, Germany\\
$^2$ Scuola Normale Superiore, Piazza dei Cavalieri 7, 56126 Pisa, Italy\\
$^3$ Institute of Physics, E\"otvos University, Pazmany P. s. 1/A, 1117 Budapest, Hungary \\
$^4$ Osservatorio Astrofisico di Roma, Via Frascati, 3300040 Monte Porzio Catone, Italia }

\date{\today}
\pagerange{\pageref{firstpage}--\pageref{lastpage}} \pubyear{2008}
\begin{document}

\maketitle
\label{firstpage}

\begin{abstract}
We propose the Apparent Shrinking Criterion (ASC) to interpret the spatial extent, 
$R_w$, of transmitted flux windows in the absorption spectra of high-$z$ quasars. The ASC can
discriminate between the two regimes in which $R_w$ corresponds either to the physical size, 
$R_{\rm HII}$, of the quasar \HII region, or to the distance, $R^{max}_w$, at which the transmitted
flux drops to $=0.1$ and a Gunn-Peterson (GP) trough appears.           
In the first case (\HII region (HR) regime), one can determine the IGM
mean \HI fraction, \xhi; 
in the second (proximity (PR) regime), 
the value of $R_w$ allows to measure the local photoionization rate and the local enhancement of the 
photoionization rate, $\Gamma_G$,  due to nearby/intervening galaxies.  The ASC has been tested against 
radiative transfer+SPH (Smoothed Particle Hydrodynamics) numerical simulations, and applied to a sample
of 15 high-$z$ ($z>5.8$) quasar spectra.
All sample quasars are found to be in the PR regime; hence, their observed spectral properties 
(inner flux profile, extent of transmission window) cannot reliably constrain the value of \xhi. Four sample 
quasars show evidence for a local enhancement (up to 50\%) in the local photoionization rate possibly produced 
by a galaxy overdensity. We discuss the possible interpretations and uncertainties of this result. 
\end{abstract}


\section{Introduction}
All high-redshift quasar spectra with a complete GP trough
(Gunn \& Peterson 1965) exhibit transmitted flux in an extended
region between the quasar redshift and the red side of the GP
trough (Fan \etal 2006, F06). These transmission windows provide a
powerful, and so far unique, probe of the physical state of the InterGalactic Medium
(IGM) beyond redshift $z>5.7$, where the onset of the GP trough erases the
information encoded in the \lya forest.  

The detection of transmitted flux at such high redshift provides a solid evidence for a reduced neutral hydrogen fraction (\xhi) within distances of 
several physical Mpc\footnote{Unless otherwise stated, distances are given in physical units.} from the quasar (F06). Such \HI decrease 
can be attributed to the locally enhanced photoionization rate produced by the quasar ultraviolet radiation. However, the 
physical interpretation of such transmission windows in terms of the IGM properties is not straightforward.  
Are these transmission windows the spectral counterparts of quasar {\it \HII regions} (HR), with the red side of the GP trough 
corresponding to the location of the quasar ionization front? Or are they instead {\it proximity regions} (PR) resulting from the 
reduced optical depth in the vicinity of a quasar whose \HII region extends further into the IGM ?
We will refer to these two interpretations as the HR and PR regime, respectively. 
It has been demonstrated that both regimes can be used to reproduce the observed spectra, depending on the assumed 
quasar luminosity and lifetime, and the ionization and thermal state of the surrounding IGM (Bolton \& Haehnelt
2007a; Maselli \etal 2007 [M07];  Wyithe, Bolton \& Haehnelt 2008 [W08]).

Previous studies have shown that under the assumption of the HR
regime, the relatively small extent of the observed transmission
windows detected in current data would imply a significant neutral
hydrogen fraction still present at 
$z\approx 6$, , that is ${x}_{\rm HI}>0.1$ (Wyithe \& Loeb 2004;
Wyithe, Loeb \& Carilli 2005; F06). 
These results are supported by independent analysis which do not explicitly assume the HR regime, based either on fitting 
the probability distribution function (PDF) of the \lya optical depth, $\tau$, within the transmission windows (Mesinger \& Haiman 2007), 
or on the comparison of \lya and Ly$\beta$ absorption spectra (Mesinger \& Haiman 2004).  
However, it is difficult to reconcile these results with the analysis of dark gaps in quasars/gamma-ray 
bursts absorption spectra (Gallerani \etal 2008a, 2008b), as well as with other independent probes of the hydrogen reionization history, as 
the Lyman Alpha Emitters surveys (\eg Malhotra \& Rhoads 2006; Dawson \etal 2007), and the integrated electron scattering optical depth measured by 
WMAP5 data release (Komatsu \etal 2009).
All these studies suggest \xhi$\ll 1$ at $z\approx 6$ and are consistent with the detection of transmitted flux in high-redshift
quasar spectra provided that the PR interpretation of the transmission windows applies. In this regime, the spectral extent of the 
windows is set by the onset of the GP trough {\it within} the quasar \HII region. In M07 this effect has been termed {\it apparent 
shrinking}, to indicate the possibility that the physical extent of the transmitting window, $R_w$, does not correspond to the 
actual \HII region size, $R_{\rm HII}$, and in fact underestimates it.
 
The condition $R_w < R_{\rm HII}$ is likely to arise from the combination of two factors: (i) the average IGM density gets high
enough to produce complete \lya absorption in spite of the high gas ionization level, and (ii) the quasar  photo-ionization rate 
drops below the minimum value required to preserve IGM transparency inside the quasar \HII region.  
  
To discriminate between the HR and PR regimes we propose here a simple (but yet solid) criterion, the 
Apparent Shrinking Criterion (ASC). Such method, in addition, allows to determine ${x}_{\rm HI}(z)$ in the HR regime, 
or to quantify local deviations from the mean cosmic photoionization rate along the line of sight (LOS) to the quasar if the PR regime holds.

\section{The Apparent Shrinking Criterion}
In this Section we introduce the basic idea underlying the {\it Apparent Shrinking Criterion}, 
whose main use is to discriminate between the HR and PR scenarios described above by using a given 
high-redshift quasar spectrum.
The ASC is based on the comparison among three spatial scales: (a) the physical size of the spectral 
transmission window, $R_w$, (b) the size of the quasar \HII region, $R_{\rm HII}$, and (c) the maximum 
value that $R_w$ can take, $R_w^{max}$. These are discussed separately in the following.  
\subsection{Size of the transmission window} 
$R_w$ is a directly measurable quantity in high-$z$ quasar spectra and it denotes the physical extent of the spectral region close 
to the quasar in which a non-zero flux can be detected. The determination of this quantity is affected by uncertainties on quasar redshift, $z_Q$, and on the flux detection limit of the instrument. An additional complication is represented by
the fact that the IGM transmissivity at $z\approx6$ is a mixture of dark gaps and transmission peaks, which makes the identification of the red side boundary of the GP trough
uncertain (e.g. isolated peaks in the translucent IGM can merge with the transmission window region associated to the quasar).
Here we adopt the operational definition given in F06, which states that \rw~ represents the physical distance from the quasar at which, 
when smoothed by a top-hat filter of 20~\AA, the observed flux drops below a transmission threshold of 0.1 (corresponding to a \lya 
optical depth $\tau=2.3$). 
\subsection{Size of the quasar \HII region}
$R_{\rm HII}$ denotes the mean radius of the quasar \HII region.  Due to the gas inhomogeneities and to radiative transfer 
effects, 
quasar \HII regions deviate significantly (up to 15\%, e.g. M07) from spherical symmetry with 
a resulting variation of the \HII region radius along different LOS (see for example Fig.~\ref{LOS}). Previous studies including RT calculations
have shown that the radius of a quasar \HII region expanding in a homogeneous IGM approximates well the actual mean value of $R_{\rm HII}$ along different 
lines of sight in the inhomogeneous case.
Hence, we use the following expression to determine the average HII region size:
\beq
\label{rHII}
R_{\rm HII}\approx 3.3
\left(\frac{\dot{N}_{\gamma}}{10^{57} {\rm s}^{-1}}\frac{t_{Q}}{10^7
{\rm yr}}\right)^{1/3}
\left(\frac{1+z}{7}\right)^{-1} x_{\rm HI}^{-1/3} {\rm Mpc},
\eeq
The above expression is valid under the additional assumption that the quasar lifetime,
$t_Q\approx 10^{7-8}$~yr, is shorter than the gas (volume-averaged) recombination timescale, 
\beq
t_{rec}(\nHI) = 2.6~ C^{-1} \left(1+z\over 7\right)^{-3}~{\rm Gyr};
\label{trecn}
\eeq
where the clumping factor $C \equiv \langle n^2\rangle/\langle n\rangle ^2 > 1$
is meant to include the effects of density inhomogeneities inside the
ionized region; we have used a helium to hydrogen number ratio $y=0.08$.
Due to its low density, typical IGM recombination times are long, and exceed $t_Q$.
\subsection{Maximum size of the transmission window}
A third fundamental spatial scale entering the formulation of the ASC is set by the monotonic dependence of the mean transmitted 
flux, $F\equiv e^{-\tau_{eff}}$, on the distance from the quasar. The detection of the GP trough in all spectra of quasars with 
$z>5.7$ implies that at these redshifts the mean IGM opacity grows large enough to completely suppress transmission; due to instrumental 
sensitivity the detection of GP trough implies an upper limit on the mean transmitted flux of  $F < 0.1$. (see \eg F06).
As a result of their large ionizing power, in the vicinity of extremely luminous quasars the IGM opacity is suppressed with respect to average and
some flux shortwards of the \lya line can be detected. Due to the geometrical dilution of the quasar ionizing flux, 
$F$ is a monotonically decreasing function of the distance from the quasar, $R$, and it will eventually fall below the detection threshold $F=0.1$;
at even larger distances, the quasar photoionization rate drops below the one produced by the background sources (QSOs and galaxies).
Given the quasar ionizing photon rate, $\dot{N}_\gamma$, and the mean intensity of the ionizing background (or equivalently, \xhi), one can 
define a radius $R^{max}_w$, as {\it the distance from the quasar such that $F(R^{max}_w)=0.1$}, under the hypothesis 
that the quasar flux extends to infinity. Stated differently, \rmax~ represents the maximum spatial extent of the 
transmission window $R_w$ associated to a given quasar. 

How can we estimate $R^{max}_w$? Analogously to $R_{\rm HII}$ in eq.~\ref{rHII}, \rmax is calculated in the 
ideal case of a uniform IGM; deviations might occur due to fluctuations along different LOS, 
as well as to uncertainties in the modeling of the IGM and quasar properties. These effects will be quantified in the next Section. 
 
For a uniform IGM the mean optical depth can be expressed as a function of redshift and of the neutral hydrogen density as  follows:
\be
\label{tau}
\tau = 4.25 \times 10^5 h^{-1} {x}_{\rm HI} \Delta
\left(\frac{\Omega_m}{0.26}\right)^{-0.5}
\left(\frac{\Omega_b h^2}{0.0241}\right) 
\left(\frac{1+z}{7}\right)^{3/2},
\ee
where $\Delta \equiv \rho/\langle{\rho}\rangle$ denotes the gas overdensity. The mean transmitted flux is however sensitive to the probability 
distribution function (PDF) of the density field, $P(\Delta)$, and can be estimated as:
\be
\label{meanFlux}
F &=& \int P(\Delta) e^{-\tau(\Delta)} d\Delta.
\ee
As quasars typically reside in biased (i.e. high-$\sigma$ peaks) regions of the cosmic density distribution,
$P(\Delta)$ might differ in their vicinity from the one describing the general IGM. Under these conditions 
we do expect in particular that the peak of the PDF is shifted towards larger $\Delta$ values. In the following, 
though, we will neglect this bias, as both theoretical models (Barkana 2002; Wang \etal 2009) and observations 
(Guimar\~aes \etal 2007) find the ambient overdensity to be significant only within a distance 
($\approx 1$~Mpc) much smaller than the observed values of $R_w$: the minimum value measured to date is the one 
in the quasar J1623+3112 at $z=6.22$ with $R_w=3.6$~Mpc.

In general, $F$ is a function of the distance $R$ from the quasar, through the dependence of $\tau$ on
\xhi, as in eq. \ref{tau}. The neutral fraction, in turn, can be derived from the condition of
photoionization equilibrium at any given $\Delta$ once the local photoionization rate $\Gamma(R)$ is assigned.
Close to the quasar, $\Gamma(R) \approx  \Gamma_Q(R)=\dot{N}_\gamma\bar{\sigma}_H /4 \pi R^2$, with
$\bar{\sigma}_H$ denoting the frequency-averaged photoionization hydrogen cross-section; furthermore 
within distances $R\approx R_{\rm HII}$, $\Gamma(R) \gg \Gamma_B$, the mean value in the IGM.
The expression for $\Gamma_Q$ given above neglects the flux suppression due to residual \HI in the ionized region; 
as discussed later on, this is a good approximation supported by the results from full 3D cosmological 
radiative transfer simulations. 
An additional contribution, $\Gamma_G$, to the total photoionization rate along the LOS can come from galaxies intervening along the line 
of sight. 

Then the local photoionization rate can be expressed as:
\beq
\label{gamma}
\Gamma(R)= \Gamma_B + \Gamma_G (R) +\Gamma_Q(R).
\eeq 
We initially neglect $\Gamma_G$ and assume that $\Gamma_Q(R)$ extends to infinity (\ie with no truncation at the \HII region 
ionization front whose location is unknown), and discuss the implications of such assumption in Sec.~4.  

In summary, the estimate of the \rmax value associated to a given quasar requires an {\it a priori} knowledge of 
$\dot{N}_\gamma$, $\Gamma_B$ and of the mean IGM temperature, $T$, which enters the determination of \xhi~ (and $\tau$, see
eq. \ref{tau}) via the \HII recombination rate. We discuss the choice of these quantities and the associated 
uncertainties in Sec.~3.  Once these quantities are fixed, \rmax~ can be estimated from photoionization equilibrium, 
and by requiring that $F (R^{max}_w) = 0.1$. From the analysis of RT calculations, it appears that 
photoionization equilibrium is a good approximation in the cases of interest. 
Note that, \rmax is univocally determined by such procedure, being $\tau(R)$ a monotonically increasing function of $R$ as a 
consequence of the geometrical dilution of the quasar ionizing flux.

\subsection{Formulating the ASC} 
Now that we have all the necessary definitions at hand, we can describe the formulation of the ASC. 
As anticipated above, the ASC is based on the comparison between 
$R_{\rm HII}$ and $R^{max}_w$.

The first case, $R_{\rm HII}>R^{max}_w$, corresponds to the PR scenario, i.e. a situation in which the  
the quasar photoionization rate is too low to keep the gas sufficiently transparent within its own \HII region.
Under these conditions, the measured size of the transmission window will be equal to its theoretically determined 
value, $R_w = R^{max}_w$, within uncertainties. 
It has already been shown that this situation is likely to occur in an ionized universe 
where quasar \HII regions can expand several tens of Mpc into the IGM (Bolton \& Haehnelt 2007; M07), albeit it
could also be found in an almost neutral gas provided that the quasar luminosity and lifetime are large enough to 
power a rapid growth of $R_{\rm HII}$. 

If instead $R_{\rm HII}< R^{max}_w$ (HR regime), the extent of the transmission window is set by the size of the 
\HII region, \ie by the sharp \xhi~ discontinuity at the location of the ionization front. In this case 
$R_w=R_{\rm HII}$: by construction, \rmax~ represents the upper limit to $R_w$ obtained by assuming that quasar 
ionizing flux extends to infinity.  The HR regime usually occurs when \xhi~ is large (\ie small $R_{\rm
 HII}$) and the quasar ionizing flux is suppressed abruptly while still dominant over the Ultra Violet Background (UVB). 
If the HR regime holds, the measured value $R_w$ gives an estimate of \xhi~ via eq.~\ref{rHII} as first proposed 
by Wyithe \& Loeb (2004).  It is worth stressing that, by their own operational definition, the PR and HR regimes 
are not associated to a particular phase of the reionization process. 
  
Note that, according to our definition, it is always $R^{max}_w \ge R_w$, within errors. However,  
\rmax is determined under the assumption that $\Gamma_G=0$, \ie neglecting a possible contribution to $\Gamma$
by sources other than the quasar and the UVB. This might well not be the case.  
However, it is easy to isolate the effects of contributing sources (most likely galaxies) along the LOS
which would be signaled by a larger than expected value $R_w > R^{max}_w$. In this case one would conclude that 
$\Gamma_G \neq 0$; in addition, it will be also possible to obtain a quantitative estimate of such contribution 
by using eqs. \ref{tau}$\div$\ref{gamma} and further imposing that $F(R_w)=0.1$.

The application of the ASC will allow to decide between the following
possibilities: if $R_w \ge R^{max}_w$ the PR regime applies and no constraints can be put on \xhi; 
if moreover $R_w > R^{max}_w$ this implies a non-vanishing contribution of galaxies to the local 
photoionization rate, $\Gamma_G \neq 0$. If instead $R_w < R^{max}_w$, the HR regime applies, $R_{\rm HII} 
\approx R_w$, and \xhi~ can be constrained. 

\section{The error budget}
In this Section we discuss and quantify the various uncertainties involved in the determination of the three characteristic 
scales entering the ASC, together with their associated error sources. 

\subsection{$R_{\rm HII}$}
$R_{\rm HII}$ denotes the average extent of the quasar \HII region and consequently the uncertainty on this quantity 
is associated to fluctuations in the density field and radiative transfer effects (\eg shadowing, filtering and 
self-shielding) which distort the shape of the quasar \HII region; the extent of the \HII region is then 
direction-dependent. The amplitude of this effect can be estimated only with the support of 3D numerical simulations. 

We have then performed a set of dedicated 3D radiative transfer simulations which follow the expansion of the \HII region in a cosmological density field
for different quasars and IGM parameters, as described in more details in Sec. 4. 
We use the outputs of such simulations to quantify the relative error $\Delta R_{\rm HII}/R_{\rm HII}$.  
For each simulation we have drawn 1000 LOS  towards the quasar and on each LOS we have estimated $R_{\rm HII}$ as the 
distance $R$ at which $\partial x_{\rm HI}/\partial R > 100$.   
$\Delta R_{\rm HII}$ is quantified as the 1-$\sigma$ dispersion of the $R_{\rm HII}$ distribution derived as above. 

Tab.~1 reports $R_{\rm HII}$ and the associated error for various simulation runs\footnote{For runs R2 and R4 these values are not given as these cases correspond to an initially highly ionized IGM in which the 
quasar \HII region becomes larger than the computational volume.}. 
We find the relative error on $R_{\rm HII}$ to be confined in the
extremely narrow range $0.07-0.11$, despite the wide range covered by
the average extents of the \HII regions which result from spanning
over the values allowed for the simulation parameters ($\dot{N}_\gamma$, $x_{\rm HI}$, $t_Q$).

\subsection{ $R_w$}
The extent of the transmission window is an observable quantity in the spectra of high-$z$ quasars whose estimate is derived 
from the measurement of the quasar redshift ($z_Q$) and the onset of the GP trough ($z_{w}$) along a given 
LOS.
The accuracy of the quasar redshift depends on the line indicators available. The most accurate determination relies on 
CO rotational transition lines from the host galaxy. Only for two of the high-$z$ quasars analyzed (see Sec.~5)
CO detection is currently available: J1148+5251 (Bertoldi \etal 2003) and J0927+2001 (Carilli \etal 2007); for these cases redshifts are 
measured with an uncertainty  $\Delta z_Q  < 0.0001$.
The next accurate indicator for systemic quasars redshift is the \MgII line which, despite being a broad line, allows accuracy of the order 
of $0.002 < \Delta z_Q < 0.004$ (Richards \etal 2002). However the \MgII transition line has a restframe wavelength of 2800 \AA, so that at $z\approx 6$ it is shifted 
to roughly 2 $\mu$m, thus requiring infrared spectroscopy which is not always available. When neither CO and \MgII 
detections are available the redshift is measured via the \lya line which has an associated error of the order of $\Delta z_Q \approx 0.02$.
In the quasars list reported in Table~1, the redshifts are tagged with a letter specifying the line indicator used. 

Extracting an accurate measurement of  $z_w$ from the observed spectra is also not trivial, as already discussed in Sec.~2.1. As mentioned there, we adopt 
the operational definition given in F06, \ie $z_w$ is taken as the redshift corresponding to the first pixel,  moving away 
from the quasar location, at which the flux smoothed by a top-hat filter of 20~\AA~drops below the threshold $F=0.1$. 
At $z\approx6$, 20~\AA~correspond roughly to $\Delta z_w \approx 0.02$. We take the latter as a measure of the error 
on $z_w$.

Aside from the intrinsic error associated to the \rw~measurement along the observed LOS, the uncertainty on $R_w$ must take into 
account the dispersion along different lines of sight toward the same target quasar. Analogously to  the case of $R_{\rm HII}$, quantifying such an error 
requires the support of 3D numerical simulations. We have used the same sample of LOS used for the $\Delta R_{\rm HII}$ determination, to extract 
the mock spectrum and the associated value of $R_w$ for each LOS (the details of such procedure are given in Sec.~4). The average \rw~values found
for the various simulations are given in Tab.~1 together with the associated 1-$\sigma$ dispersion of the sample.  As already pointed out in
our previous work (M07),  $R_w$ fluctuations along different LOS are significantly larger with respect to the $R_{\rm HII}$ ones: this is due to 
the fact that \rw~depends on the \lya optical depth which is extremely sensitive to tiny fluctuations of \xhi, while 
$R_{\rm HII}$ is set by the sharp \xhi ~transition occurring at the ionization front.
Differently from $R_{\rm HII}$, \rw ~has a stronger dependence on the physical state of the gas in which the
quasar is embedded. We find $\Delta R_w / R_w$ values ranging from 0.17 to 0.58, with the general trend of $\Delta R_w / R_w$ increasing with decreasing values of the ratio
$\dot{N}_\gamma /  \Gamma_B$. This behavior is determined by the fact that $R_w$ gets smaller for smaller $\dot{N}_\gamma$, 
while for higher $\Gamma_B$ values the dispersion, $\Delta R_w$, increases as a consequence of a more sensitive response
to density fluctuations. Given the limited a-priori knowledge of $\dot{N}_\gamma /  \Gamma_B$ in this study 
we assume the average value $\langle \Delta R_w / R_w \rangle = 0.3$. 

\subsection{$R_w^{max}$}
As described in Sec.~2.3, the determination of  \rmax involves the assumption of the following parameters: the quasar intrinsic luminosity, $\dot{N}_\gamma$, 
the gas temperature, $T$, and the UVB photoionization rate, $\Gamma_B$.

Quasars luminosities are commonly estimated by converting the observed magnitudes in number of ionizing photons 
emitted per unit time. Such a procedure requires the assumption of a specific template for the quasar intrinsic spectrum, 
which is poorly constrained for high redshift quasars. 
In principle each quasar is characterized by its own spectral energy distribution and it would be possible to associate
at each observed object a power law with a specific spectral index inferred from infrared data, which however are not always available for the current 
analysis.
In the current work we derive $\dot{N}_\gamma$ by using the measured absolute magnitude $M_{1450}$ at 1450 \AA,  and 
assuming the spectral template by Telfer \etal (2002). This is the composite spectrum of lower redshift 
quasars ($z \lesssim 3.7$),  which is well fitted by a double power-law with a break close to the \lya line at $\lambda=1280$ \AA~ 
and spectral indexes $\alpha_{EUV}=-1.76$ and $\alpha_{NUV}=-0.69$.
As done in Yu \& Lu (2005), we take into account the uncertainty associated to the template by considering also the Telfer double power-low
fits to the subsamples of radio-quiet and radio-loud quasars, which has spectral indexes ($\alpha^{rl}_{EUV}=-1.96$, $\alpha^{rl}_{NUV}=-0.67$)
and ($\alpha^{rq}_{EUV}=-1.57$, $\alpha^{rq}_{NUV}=-0.72$) respectively. 

The IGM temperature as well as the UVB photoionization rate, $\Gamma_B$, are also poorly constrained at high redshift. 
Nevertheless the statistical analysis of the \lya forest detected in high-$z$ quasar spectra, as well as both analytic and numerical theoretical models of the reionization process, 
allow to bracket the above quantities into a relatively narrow range of possible values (\eg Choudhury \& Ferrara 2006; Gallerani \etal 2008a).
In the application of the ASC to the observed high-$z$ quasar spectra presented in this paper and discussed in detail in Sec.~5, we assume $\Gamma_{B}$ 
to evolve according to the Early Reionization Model (ERM) developed by Gallerani \etal (2008a). 
In such model, reionization completes at $z_{rei}\approx 7$ after an extended partial 
ionization phase, necessary to recover the recent determination of the electron scattering optical depth found 
from WMAP5 data (Dunkley \etal 2009). Such model also simultaneously accounts for most of the available data 
including Ly$\alpha$/Ly$\beta$ GP opacity, Lyman Limit Systems, cosmic
star formation history, the number density of 
high-redshift sources, and the Luminosity Functions of Lyman Alpha Emitters at $4.5<z<6.6$ (Dayal \etal 2008). 
Moreover, the UVB photoionization rate predicted by the ERM is consistent with the results by Bolton \etal 
(2005, 2007) in the range $4<z<6$. 
At $z\approx 6$ the latter authors provide a solid upper limit $\Gamma_B \le 0.34 \times 10^{-12}$~s$^{-1}$,
which we take here to quantify the uncertainties on $\Gamma_B(z)$. 
Smaller values of $\Gamma_B$ do not affect significantly the \rmax estimate, due to the fact that
at the characteristic distances involved (several physical Mpc) the local photoionization rate is still dominated by the quasar ionizing flux.

The ERM prescribes IGM temperatures in the range $T =(1\pm 0.5)\times 10^4$~K.  
We take $T =(1.5\pm 0.5)\times 10^4$~K, to allow for a possible local boost induced by the quasar.  
Note in fact that, as seen from our RT simulations presented in Fig. 1, the extra heating provided by the quasar 
can rise the gas temperature above $2\times10^4$~K only for the case in which the quasar ionization 
front expands in a completely neutral gas, a possibility already excluded by various high-$z$ experiments (see Introduction).

\begin{figure*}
\centerline{\psfig{figure=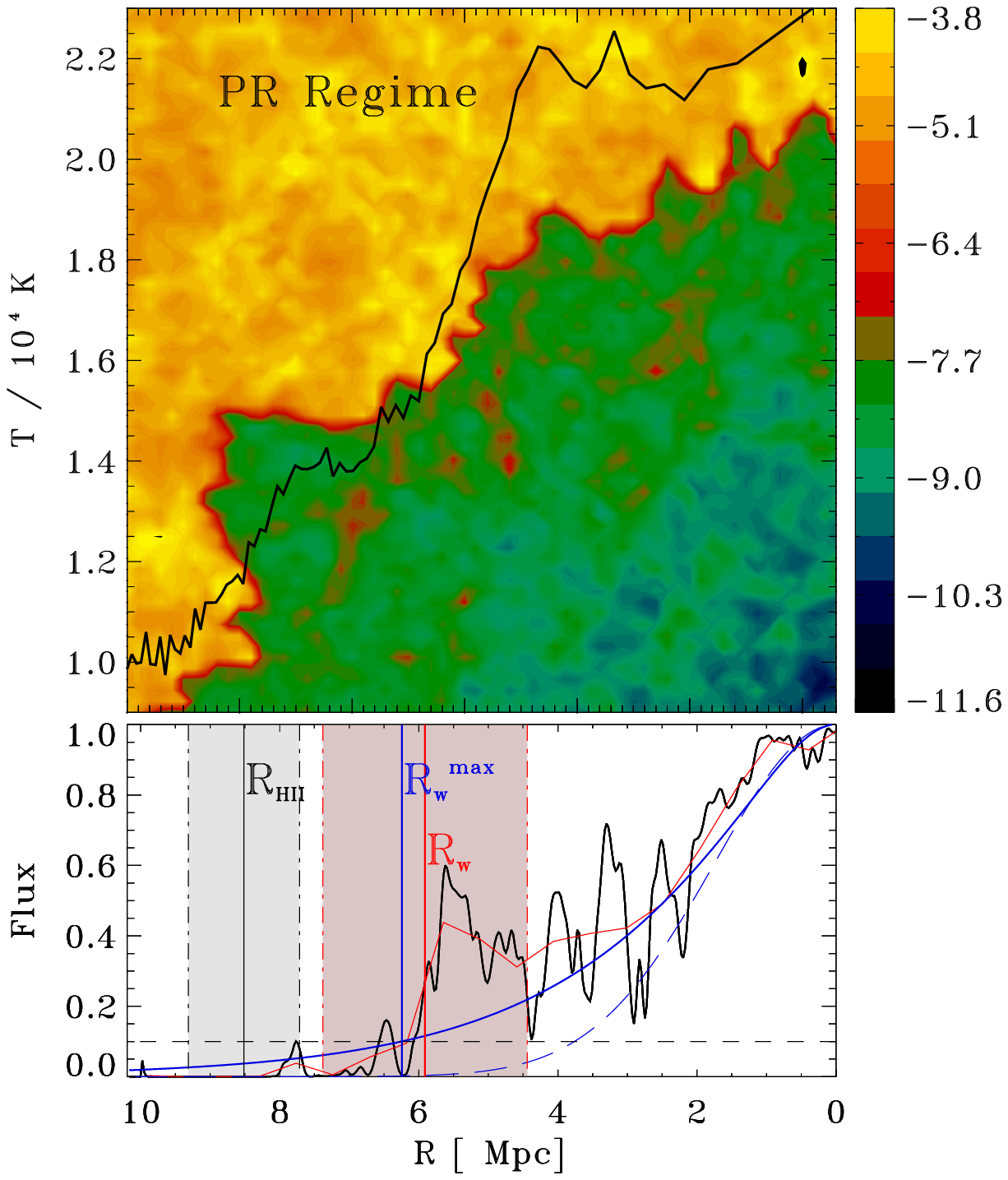,height=10.5cm}\psfig{figure=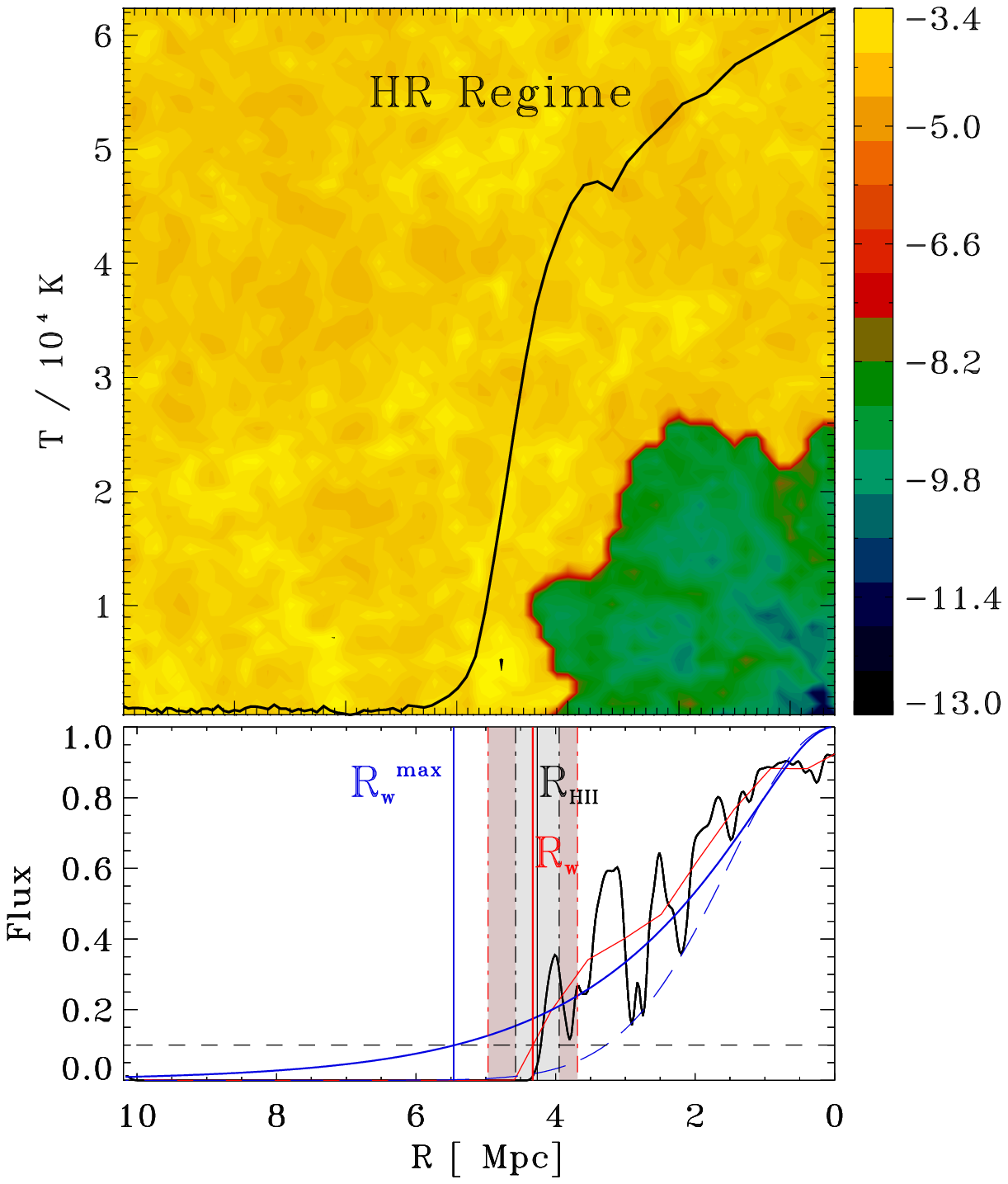,height=10.5cm}}
\caption{Simulated neutral hydrogen number density maps (upper panels, color bars
give $\log n_{\rm HI}$ values) and corresponding mock
spectra (lower) along a typical LOS through the quasar for
the  PR (left, run R3) regime and HR (right, run R5). The maps are 10.1 physical Mpc
on a side with the quasar located in the right-bottom
corner. Overplotted are the corresponding mean temperature profiles.
The lower panels show the full-resolution spectrum  (black solid line),
the same smoothed at 20~\AA (red solid), the mean transmitted flux
derived from eq. \ref{meanFlux} (solid blue) and using
$F=e^{-\tau(\Delta=1)}$ (dashed blue). 
The values of $R_w$ (measured along the plotted spectrum) and of 
$R_w^{max}$ and  $R_{\rm HII}$ (averaged over 1000 LOS), are shown
by the vertical lines, along with the associated 1-$\sigma$ error (shaded band). 
}
\label{LOS}
\end{figure*}

\begin{table*}
\centering
\caption{Tests of the ASC with simulations}
\begin{tabular}{|c|c|c|c|c|c|c|c|c|c|}
\hline
Run & $\dot{N}_{\gamma,57}$ & ${x}_{\rm HI}$ & $\Gamma_B$ & $t_Q$ & $R^{max}_w$ & $R_w$ & $R_{\rm HII}$ & Regime  \\ 
                  &  &    & [$10^{-13} s^{-1}$] &  [$10^7$ yr] &  [Mpc] & [Mpc] & [Mpc] & \\ 
\hline
\hline
R1 & $0.56$ &$0.1$& $1.25\times 10^{-3}$  & $1$ &$2.51 (2.96)$ & $2.74 \pm 0.96$ &  $5.32\pm0.61$ & PR\\ 
R2 & $0.56$ & $10^{-4.2}$ & $0.8$& $1$ & $2.30 (2.43)$ & $2.74 \pm 1.60$ &----& PR \\ 
R3 & $2$ & $0.1$ & $1.25\times 10^{-3}$& $1$ & $4.83 (5.35)$ & $5.50 \pm 1.47$ & $8.51\pm0.80$ & PR \\ 	
R4 & $2$ & $10^{-4.2}$ & $0.8$& $1$ & $4.51 (4.93)$ & $5.66 \pm 1.86$ & ----& PR \\ 
R5 & $2$ & $1$ & $0.0$& $1$ & $3.4 (4.7)$ & $3.70 \pm 0.64$ & $4.26\pm0.31$ & HR \\ 
R6 & $2$ & $1$ & $0.0$& $2$ & $4.2 (5.7)$ & $4.55 \pm 0.89$ & $5.39 \pm 0.42$ & HR \\ \hline
\end{tabular}
\end{table*}
\section{Testing the ASC with simulations}
Before applying the ASC to observations we have tested it through a set of numerical simulations which combine SPH 
(Springel \& Hernquist 2003) and full 3D radiative transfer (RT)
calculations performed with {\small CRASH} (Maselli, 
Ferrara \& Ciardi 2003).  These tests are meant to check the validity of the ASC and more precisely to verify that the semi-analytic 
method devised to determine \rmax gives reliable results. If so, when applied to the synthetic spectra extracted from the 
simulations, the ASC should be able to correctly predict the physical regime (PR or HR) prevailing in the computational 
volume.

The simulations setting is similar to the one described in M07, to which we defer the reader for details. The two main 
differences here are the inclusion of helium physics and the newly developed ``colored packets" RT algorithm (Maselli, 
Ciardi \& Kanekar 2008) which allows to more reliably compute the effects of hard spectrum sources.
This allows to extract mock spectra with improved precision. 
Using the SPH density field at $z=6.093$ as an input, we have run
{\small CRASH} for different combinations of the free 
parameters ($\dot{N}_\gamma$, $t_Q$, ${x}_{\rm HI}$). The set of simulations performed is listed in Table~1, 
where we give the free parameters adopted for each run, identified by the name in first column, along with values of $R^{max}_w$, $R_w$, 
their associated  1-$\sigma$ errors and the corresponding regime determined from the simulations.
From each simulation we build synthetic \lya absorption spectra along 1000 LOS through the box, smooth the spectra at 20~\AA as
described in M07, and derive the average value of $R_w$. 
The corresponding $R^{max}_w$ is evaluated following the procedure described in Sec~2.3 adopting the values of 
$\Gamma_B$ and $\dot{N}_\gamma$ of the specific RT run, 
as well as the numerically computed/LOS-averaged neutral hydrogen density and temperature profiles. 
Due to the inhomogeneities of the density field and to RT effects, the quasar \HII regions found in our numerical 
simulations are not spherical, with significant scatter of the \HII region size along different directions (\eg see 
Fig.~\ref{LOS}). We then compute $R_{\rm HII}$ by  averaging the value over all the LOS. For all the simulated 
configurations in which the \HII region is confined inside the computational volume (R1, R3, R5, R6), we find that the average values
of $R_{\rm HII}$ are consistent with the approximated values given by eq.~\ref{rHII} within few percent. This result shows 
photoionization equilibrium, assumed in the calculation of \rmax (Sec. 2.3), to be an excellent approximation.

Two examples of PR (left) and HR (right) regimes are shown in Fig.~\ref{LOS}, taken from the runs R3 and R5, respectively.  
The maps show \HI number density cuts in a region of 50 $h^{-1}$ Mpc (comoving) on a side with the quasar
located in the right-bottom corner. Overplotted in the maps is the averaged temperature profile within the same region. 
In the corresponding bottom panels of the same Figure, the black (red) solid lines show the full-resolution (20 \AA~ smoothed) absorption 
spectrum  along a random LOS to the quasar, from which we determine $R_w$ according to the operational definition  given in Sec.~2.3. 
The mean transmitted flux derived from eq. \ref{meanFlux} is shown by the solid blue lines, while the dashed curves 
represent the mean transmitted flux for the case of an homogeneous density field at the mean density, \ie  $P(\Delta)=\delta(1)$.  

By comparing the analytical transmitted fluxes curves to the mock spectra, it is evident that to correctly 
compute the IGM mean opacity at high-$z$, the proper PDF of the density field, and particularly of underdense regions, must be 
taken into account.  At high redshift, in fact, $\tau(\Delta=1)$ grossly overestimates the effective opacity. 
This occurs because fluctuations above a certain overdensity threshold, $\Delta_{th}$, start to become opaque due to the progressive increase 
of the average cosmic density with $z$, which prevents the gas at $\Delta \ge \Delta_{th}$ to remain sufficiently ionized to be transparent. 
The value of $\Delta_{th}$ decreases gradually with redshift and becomes $< 1$ at $z\approx 6$.

Fig. 1 illustrates visually the relation among $R_w$, $R^{max}_w$ and $R_{\rm HII}$ as a function of the different
physical conditions of the gas in the two regimes. 
Note that in these plots the mock spectra as well as \rw~are computed along a single random LOS;  
\rmax is determined analytically, and $R_{\rm HII}$ is the average value over all the LOS. This 
choice is meant to reproduce the closest analog to the actual experiment. 
In the PR regime (left panel), we find $R^{max}_w \approx R_w$  as expected, implying, according to the ASC, 
that the quasar \HII region should be larger, $R_{\rm HII}> R^{max}_w \approx R_w$. This prediction is confirmed, 
as seen from the position of $R_{\rm HII}$ shown in the panel.  On the contrary, in the situation in which  
$R^{max}_w > R_w$, as in the right panel corresponding to the HR regime, $R_w$ truly measures the \HII region 
radius. Again the simulation results confirm the validity of the ASC prediction. 
In this case by inverting eq. \ref{rHII}, we obtain a (nominal) value \xhi$ = 0.89 \pm 0.2$, consistent with the 
neutral hydrogen fraction, \xhi$=1$, imposed as initial condition in this run\footnote{ 
Relativistic effects related to finite light propagation on the apparent shape of the \HII regions do not 
affect the direction along the LOS and thus the validity of the ASC which applies to single LOS.}.

More systematic tests of the ASC have been envisaged with the set of simulations reported in Tab.~1, 
which span the allowed values of the free parameters of the problem
($\dot{N}_\gamma$, $t_Q$, ${x}_{\rm HI}$). 
For each run we give here the values of $R^{max}_w$, $R_w$, and their associated 1-$\sigma$ error 
(see discussion in Sec.~3). We provide two different estimates for $R^{max}_w$: the first (fiducial) is obtained by using the average 
$n_{\rm HI}$ profiles extracted from the RT simulations, the second, $R^{max}_w(noRT)$ (in brackets) by neglecting 
RT effects. In this latter case $x_{\rm HI}$ inside the \HII region is calculated by assuming 
that the SPH simulated density field is at photoionization equilibrium with the {\it unattenuated} quasar flux plus
background: this method does not account for (i) residual \HI inside the \HII region, 
and (ii) the drop of the quasar flux beyond the ionization front. 

From the data in Tab. 1 we find that $R^{max}_w$ is within the $R_w$ range derived from the mock spectra;
$R^{max}_w(noRT)$ is in agreement with $R_w$ for PR cases, but it
exceeds $R_w$ for the R5 and R6 (HR regime) runs, as
expected from the ASC:  $R_w$ in these cases corresponds to the \HII region radius (eq. \ref{rHII}), a result 
confirmed by the analysis of these simulation outputs. 

The comparison between the fiducial and no-RT (bracketed) $R^{max}_w$ values shows furthermore that RT effects associated 
to the internal residual \HI opacity can be safely neglected in the calculation of $R_w^{max}$, whereas the 
flux drop occurring in the HR regime beyond the ionization front cannot be disregarded. 
Stated in a different manner, the power of the ASC to discriminate between PR and HR regimes 
exploits the strong flux drop-off taking place at the \HII region ionization front.
\begin{table*}
\label{table2}
\centering
\begin{tabular}{|c|c|c|c|c|c|c|c|c|c|}
\hline
${\rm Quasar}$ & $ z_Q $ &$z_w$& $M_{1450}$  & $\dot{N}_{\gamma,57}$ & $\dot{N}_{\gamma,57}$ & $\dot{N}_{\gamma,57}$ & $R_w$ & $R_w^{max}$ \\ 
&&&&Fiducial& Radio Loud & Radio Quiet && & \\ \hline
\hline
$J0927+2001$ &  $5.77^a$ & $5.700^b$ &$-26.78$& $1.00$ &  $0.85$ &$1.15$  & $4.96$  & $4.45$\\ 
$J0002+2550$ &  $5.80^b$ &  $5.650^b$ &$-27.70$& $2.33$ &  $1.98$&  $2.68$ &  $10.68$ & $7.15$ \\
$J0836+0054$ &  $5.81^c$ & $5.721^d$ &$-27.83$& $2.63$ &  $3.02$&  $2.23$ &  $6.24$ & $7.55$ \\
$J1436+5007$ &  $5.83^b$ & $5.720^b$ &$-26.51$&  $0.79$ &  $0.90$&  $0.66$ &  $7.69$ & $3.75$ \\
$J0005-0006$ &  $5.85^d$ & $5.771^d$ &$-26.42$& $0.72$  &  $0.82$&  $0.61$ &  $5.45$ & $3.55$ \\
$J1411+1217$ &  $5.93^c$ & $5.823^d$ &$-26.70$& $0.93$ &  $1.07$&  $0.80$ &  $7.20$ & $3.85$ \\
$J0818+1722$ &  $6.00^b$ & $5.920^b$ &$-27.37$& $1.72$ &  $1.98$&  $1.46$ &  $5.22$ & $4.95$ \\
$J1306+0356$ &  $6.01^d$ & $5.916^d$  &$-27.14$& $1.39$ &  $1.60$&  $1.18$ &  $6.13$ & $4.45$ \\
$J1137+3549$ &  $6.01^b$ & $5.910^b$ &$-27.08$& $1.32$ &  $1.51$&  $1.12$ &  $6.53$ & $4.25$ \\
$J1630+4012$ &  $6.05^b$ & $5.980^b$ &$-26.11$& $0.54$ &  $0.62$&  $0.46$ &  $4.48$ & $2.65$ \\
$J1602+4228$ &  $6.07^b$ &  $5.950^b$&$-26.82$& $1.04$ &  $1.19$&  $0.88$ &  $7.70$ & $3.65$ \\
$J1250+3130$ &  $6.13^b$ & $6.030^b$ &$-27.11$& $1.35$ &  $1.56$&  $1.15$ &  $6.25$ & $4.05$ \\
$J1623+3112$ &  $6.25^c$ & $6.160^f$ &$-26.67$& $0.90$ &  $1.04$&  $0.77$ &  $5.39$ & $3.05$ \\
$J1030+0524$ &  $6.31^d$ & $6.217^d$&$-27.15$&  $1.41$ &  $1.62$&  $1.20$ &  $5.52$ & $3.65$ \\
$J1148+5251$ &  $6.42^e$ &  $6.325^f$ &$-27.82$& $2.60$ &  $2.99$&  $2.22$ &  $5.37$ & $4.75$ \\ \hline
\end{tabular}

\caption{Quasars Sample. Quasar redshifts are tagged according to the spectral lines used for their measurement 
and to reference works as follows: 
(a) CO (Carilli {\tt et al.} 2007), 
(b)  Ly$\alpha$  (Fan {\tt et al.} 2006),
(c) \hbox{Mg~$\scriptstyle\rm II\ $} (Jiang {\tt et al.} 2007),
(d) \hbox{Mg~$\scriptstyle\rm II\ $} (Kurk {\tt et al.} 2007), 
(e) CO (Bertoldi {\tt et al.} 2003),
(f) (White {\tt et al.} 2003.)}

\end{table*}

\begin{figure*}
\centerline{\psfig{figure=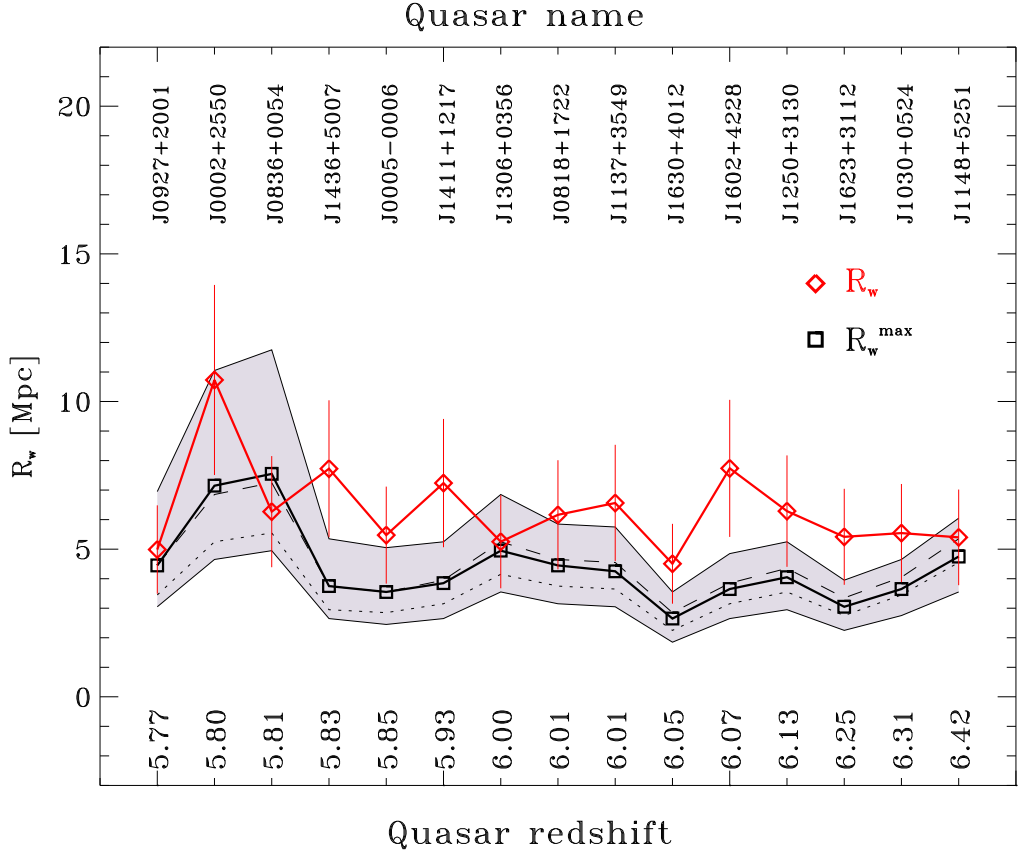,height=12.cm}}
\caption{Comparison between the observed $R_w$ and corresponding values of \rmax for the sample quasars listed in Tab. 2.
Red diamonds show the measured $R_w$, black squares are the values obtained for \rmax. The shaded region accounts for the 
errors associated to the uncertainties in the gas temperature and quasar template; the dashed (dotted) line connects 
\rmax~values obtained assuming $\Gamma_{B,-12}=0.34$ ($\Gamma_B = 0$), at fixed IGM temperature ($T=1.5\times 10^4$) 
and Telfer quasar template.}
\label{zz}
\end{figure*}

\section{Application to data}
Having successfully tested the ASC, we apply it to a sample of high redshift quasar spectra taken from F06. 
The selected quasars name, redshift, ionizing photon rate, observed $R_w$ value along with our fiducial \rmax estimate, 
are reported in Table~2.  
The redshift values are taken from F06, unless measurements of low-ionization lines like \MgII (Kurk \etal 2007; Jiang \etal 2007) or of CO rotational lines
(J0927+2001 from Carilli \etal 2007; J1148+5251 from Bertoldi \etal 2003) do exist. 
The QSO ionizing rates have been derived from their magnitudes, as discussed in Sec.~3.3 assuming the Telfer composite 
spectrum. The \rmax values are determined from the semianalytic procedure described in Sec.~3.3 and 
adopting the overdensity PDF extracted from the SPH simulation described in Sec.~4. 
In addition, $\Gamma_{B}$ and the IGM temperature, $T$, are assumed to evolve according 
to the Early Reionization Model (ERM) developed by Gallerani \etal (2008a). The details of the above models 
have been given in Sec.~3.3, where we also discuss the uncertainties associated to $\Gamma_{B}$ and $T$.

Fig. \ref{zz} shows the comparison between $R_w$ and \rmax for the quasar sample in Tab. 2. 
As discussed in Sec.~3.2, the errors on $R_w$ are dominated by the uncertainty associated to the dispersion along 
different LOS. The  \rmax values shown in Fig.~2 by the black squares are obtained by adopting an IGM temperature $T=1.5\times10^4$K. 
The shaded region represents the error on \rmax associated to the uncertainties on the temperature and on the spectral
template: the upper (lower) solid line connects the values associated with the larger (smaller) photoionization rates corresponding to the double power-law
fitting the radio-quiet (radio-loud) composite spectrum, and to the upper (lower) limit on the temperature consistent with the ERM, 
$T=2\times 10^4$ K ($T=10^4$ K). Note that higher temperatures at fixed density yield lower optical depths,  
larger transmitted fluxes and consequently higher \rmax values.
Fig. 2 also shows the \rmax values obtained for the fiducial $T$ and $\dot{N}_\gamma$ values, but with different 
assumptions on $\Gamma_B$: we consider the extreme cases of (i) a constant $\Gamma_B$ equal to the upper limit given 
by BH07 of $\Gamma_{B,-12}=0.34$ (dashed-black line), and (ii) $\Gamma_B=0$ (dotted-black line).  The small differences among these two 
extreme cases demonstrates that uncertainties in the intensity of the UVB do not sensibly affect the ASC. 
This results from the fact that, at the scales relevant to ASC, the local ionization rate is dominated 
by the quasar flux, \ie $\Gamma\simeq \Gamma_Q$.

Fig. \ref{zz} represents the central result of this paper, i.e. the application of the ASC to a sample
containing the most distant quasars, and it allows to draw a number of interesting conclusions. None of the 
analyzed quasars satisfies the condition $R_w < R^{max}_w$ (within uncertainties), indicating
that all sample QSO are in the PR regime.  As a consequence, current observations cannot provide 
reliable estimates of the IGM ionization state by studying their \HII region.  

Secondly, we find $R_w>R^{max}$ for the quasar J1602+4228 at $> 2-\sigma$ significance level, 
and for the quasars J1436+5007, J1411+1217, J1623+3112 at a lower significance level. 
According to the ASC this occurrence signals a locally enhanced photoionization rate.  
As discussed in Sec.~2.4 this could be due to either a higher-than-average number density 
of galaxies clustered around the quasar or a local enhancement of more distant galaxies along the observed LOS, 
but not directly associated to the overdensity in which the quasar is located. 
Another possibility is that the contribution to the local photoionization rate due to the galaxies around the 
quasar is increased by the enhanced mean free path produced by the quasar ionizing flux. In order to decide among 
these possibilities accurate spectroscopic redshift determinations from deep galaxy surveys of the quasars fields
would be necessary, which  unfortunately are not available yet.  

Independently of its origin, the ASC allows to directly infer the amplitude of $\Gamma_G$ at $R_w$ 
by progressively increasing $\Gamma(r)$ (eq.~\ref{gamma}) until the condition $R^{max}_w=R_{w}$ is met. 
Following this procedure we obtain: $\Gamma_G^{J1436} = 1.63  \times10^{-13}$ s$^{-1}$,
$\Gamma_G^{J1411} = 1.85 \times10^{-13}$ s$^{-1}$, $\Gamma_G^{J1602} = 2.38  \times 10^{-13}$ s$^{-1}$ and 
$\Gamma_G^{J1623} = 3.09 \times10^{-13}$ s$^{-1}$. 
These values are comparable to the average intensity of the UVB at the same redshift:  $ 0.35 < \Gamma_G/\Gamma_B < 0.85$. 
Furthermore, for all these four QSOs, we find that $\Gamma_G$ is comparable or exceeds $\Gamma_Q$ at $R_w$, with derived 
values of the ratio $\Gamma_G/\Gamma_Q = 1.25, 1.05, 1.49, 1.02$.
Such ionizing photon contribution from galaxies is crucial to explain the width of the transmission windows. 

\section{Discussion}
We have proposed the Apparent Shrinking Criterion (ASC), a simple
but yet robust semi-analytic method, based on the apparent shrinking effect (Maselli \etal 2007), 
which can be applied straightforwardly to the interpretation of transmission windows in the 
the spectra of distant quasars. 
The ASC is based on the comparison between three spatial scales associated with a quasar: 
these are $R_{\rm HII}$, the average physical radius of the \HII region, the extent of the transmission window detected in its spectrum, $R_w$,  and $R^{max}_w$, the distance 
from the quasar at which the mean transmitted flux drops to $F=0.1$ in the absence of attenuation due to 
either \HI internal to the \HII region or the ionization front of the latter.

The ASC applied to the spectra of single high-$z$ QSOs provides a novel method to discriminate among 
the two scenarios previously proposed in the literature to explain the QSO transmitted flux. 
In the first one the extent of the observed transmission window, $R_w$, is set by the sharp drop  
of the quasar ionizing flux at the ionization front location, so that $R_w$ gives a measure of
$R_{\rm HII}$. If the quasar \HII region grows large though, the unattenuated/geometrically diluted quasar flux, 
gets progressively weaker and eventually falls below the threshold necessary to keep the IGM transparent to  
\lya radiation. As a result, the observed $R_w$ saturates at the theoretically estimated upper limit $R^{max}_w$, 
independently of the actual extent of the \HII region, so that $R_w \approx R^{max}_w < R_{\rm HII}$.
We refer to the first/second case as the HII region (HR)/Proximity Region (PR) regime, respectively.  
In the HR regime, one can determine \xhi, the IGM mean \HI~fraction. If the PR regime holds instead, the value 
of $R_w$ can be used to constrain the local (i.e. at a distance $R_w$ from the quasar) intensity of the 
ionizing radiation and indirectly infer the contribution of intervening clustered galaxies to the 
local photoionization rate.  

As formulated in this paper, the ASC can only be applied to single objects and at the moment it is unclear
whether it could be extended to get information on the average IGM properties at high redshift.
At the moment, the ASC must be interpreted as a useful local test of reionization, \ie in regions 
surrounding luminous quasars. Nevertheless, by separately collecting information from a large sample of quasars 
in a given redshift range, the ASC can eventually give new constraints on the global reionization history. 
We leave the analysis of this further extension of the ASC for future work. 

The novelty introduced by the ASC is that it makes possible to assess the physical regime (HR or PR) characterizing 
an observed quasar.  Previous works have instead focused on extracting 
global information on the IGM redshift evolution from generic trends in the data, e.g.
the dependence of $R_w$ on redshift (F06; W08), statistical fits to the $R_w$
distribution function (Wyithe \& Loeb 2004), and/or fits to the flux profiles
within the transmission windows (Mesinger \& Haiman 2007). The drawback of the first two approaches is that, 
in order to carry such studies, it is necessary to make a scaling of $R_w$ to a reference quasar
luminosity, a procedure which is valid only in the case in which the quasars are all in the HR regime. 
By implicitly making this assumption the information encoded in the observed spectra is
degraded. The same happens in the flux profile fitting procedure, which is meant to constrain the most 
probable values for the triplet  ($R_{\rm HII}$, $\dot{N}_\gamma$, \xhi). Although in this case 
the hypothesis that $R_w$ identifies the \HII region radius is not made explicitly, the transmission profile shape 
is sensitive to the three parameters above only in the HR regime; in the PR regime the profiles 
are sensitive to $\dot{N}_\gamma$, but completely degenerate with $R_{\rm HII}$, and $x_{\rm HI}$.

The Apparent Shrinking Criterion has been extensively and successfully tested, by  applying it to synthetic 
absorption spectra extracted from a set of state-of-the-art numerical simulations. 
Next the ASC has been applied to a sample of high redshift quasar spectra taken from Fan \etal (2006).

The main result is that all quasars in the analyzed sample are found to be in the PR regime and hence 
their spectral observables (inner flux profile, extent of transmission windows) 
cannot be used to reliably constrain the value of the average IGM neutral hydrogen fraction. 
Nevertheless, for quasars in the PR regime the ASC offers a novel method to determine the local
photoionization rate, analogous to the proximity effect at lower redshift (Bajtlik \etal 1988). 
As an outcome of our analysis, we find that all the observed quasar spectra considered are consistent 
with the Early Reionization Model described in Gallerani \etal (2008a), which predicts reionization to 
be complete at $z \approx 7$ and thus the IGM at $z \approx 6$ to be highly ionized. 

Beside these general results, we have found that four objects (J1436+5007, J1411+1217, J1602+4228 and J1623+3112) 
show the presence of an additional contribution, $\Gamma_G \ne 0$, to the local photoionization rate.
The origin of this extra ionizing radiation is still unclear. Such flux
could be provided by (i) galaxies clustered in the biased high density regions in which quasars are expected to form, 
(ii) a galaxy overdensity intervening along the observed LOS to the quasar and not directly associated to the quasar overdensity, 
(iii) galaxies within the quasar \HII region whose contribution to the photoionization rate is boosted by the quasar-enhanced mean free path of ionizing photons. 
For the four quasars above we found that $1.02\le \Gamma_G/\Gamma_Q \le 1.49$, thus showing 
that for these cases the galaxy ionizing flux is comparable to the quasar one at $R_w$.
The interpretation of this result within structure formation scenarios might give interesting clues on the environmental 
impact of high-$z$ quasars. 

In the process of testing the ASC with our RT+SPH simulations, we have shown a further interesting point: 
a correct estimate of the IGM mean opacity at high-$z$ requires to properly weight the contribution from 
underdense regions with their actual probability distribution function. 
At high redshift in fact, fluctuations with overdensity above a certain threshold 
start to become opaque due to the progressive increase of the average cosmic density with $z$, and hence do 
not contribute to transmission even if they are still highly ionized. This overdensity threshold decreases 
gradually with redshift and it falls significantly below unity at $z\approx 6$.  

As a final remark, we emphasize that the lack of sample quasars in the \HII region regime, does not dismiss 
the potential use of quasar transmission windows to directly measure $x_{\rm HI}$ at $z>6$: in fact, we do 
expect that with the detection of higher redshift and/or lower luminous objects expected in the near future, 
spectra in the HR regime will likely be found.

\section*{Acknowledgments}
We thank G. Worseck for useful comments, X. Fan for providing some of the data listed in Tab.~2; 
useful discussions with J. Bolton and T. Choudhury are acknowledged. 
AM is supported by the DFG Priority Program 1177. SG acknowledges the support by the Hungarian National 
Office for Research and Technology (NKTH), through the Pol\'anyi Program.

\label{lastpage}
\end{document}